\documentclass[aps,prl,reprint,superscriptaddress]{revtex4-1}

\usepackage{graphicx}
\usepackage{subfigure}
\usepackage{color}
\usepackage{amsmath, amssymb}
\usepackage[linktocpage,colorlinks=true,linkcolor=blue,citecolor=blue,breaklinks=true]{hyperref}
\usepackage{verbatim}
\usepackage{breakcites}
\usepackage[version=3]{mhchem}

\begin{document}

\preprint{}

\title{Tailoring boundary geometry to optimize heat transport in turbulent convection}

\author{Srikanth Toppaladoddi}
\affiliation{Yale University, New Haven, USA}
\affiliation{Mathematical Institute, University of Oxford, Oxford, UK}

\author{Sauro Succi}
\affiliation{ Istituto per le Applicazioni del Calcolo ``Mauro Picone" (C.N.R.), Rome, Italy}

\author{John S. Wettlaufer}
\affiliation{Yale University, New Haven, USA}
\affiliation{Mathematical Institute, University of Oxford, Oxford, UK}
\affiliation{Nordita, Royal Institute of Technology and Stockholm University, SE-10691 Stockholm, Sweden}

\email[]{john.wettlaufer@yale.edu}

\date{\today}

\begin{abstract}
By tailoring the geometry of the upper boundary in turbulent Rayleigh-B\'enard convection we manipulate 
the boundary layer -- interior flow interaction, and examine the heat transport using the Lattice Boltzmann method. For fixed amplitude and varying boundary wavelength $\lambda$, we find that the exponent $\beta$ in the Nusselt-Rayleigh scaling relation, $Nu-1 \propto Ra^\beta$,  is maximized at $\lambda \equiv \lambda_{\text{max}} \approx (2 \pi)^{-1}$, but decays to the planar value in both the large ($\lambda \gg \lambda_{\text{max}}$) and small ($\lambda \ll \lambda_{\text{max}}$) wavelength limits. The changes in the exponent originate in the nature of the coupling between the boundary layer and the interior flow. We present a simple scaling argument embodying this coupling, which describes the maximal convective heat flux. 
\end{abstract}


\maketitle

Thermal and compositional convection underlie the behavior of a wide range of systems from planetary and stellar interiors and the motions of Earth's atmosphere and oceans, to the solidification of multicomponent melts \cite{kadanoff2001, Worster:2000, wettlaufer2011}. The simplest setting to study thermal convection is in a Rayleigh-B\'enard cell \cite{chandra2013}, wherein the flow is controlled by the Rayleigh number ($Ra$), which describes the ratio of buoyancy to dissipative forces, and the Prandtl number ($Pr$), which is the ratio of momentum to thermal diffusivities of the fluid, and the aspect ratio ($\Gamma$). 

The key quantity of interest is the vertical heat flux across the cell, expressed in non-dimensional form as the Nusselt number, $Nu(Ra, Pr)$, which describes the ratio of the total heat flux to the heat flux solely due to conduction. For $Ra \gg 1$, the function $Nu(Ra, Pr)$ is usually sought in the form of a scaling law: $Nu \sim Pr^{\zeta} Ra^{\beta}$. The value $\beta = 1/3$ emerges from the classical argument that when $Ra \gg 1 $ the dimensional heat flux should become independent of the depth of cell, implying that the boundary layers (BLs) at the upper and lower surfaces do not interact \cite{priestley1954, malkus1954, howard1966}. However, the exponents obtained from experiments and numerical simulations range from approximately $\beta = 2/7$ \cite{castaing1989, kerr1996, doering2009, urban2011, urban2012} to $\beta = 1/3$  \cite{urban2011, urban2012, sreenivasan2000, verzicco2003, niemela2006}. Theories with specific assumptions concerning the structure of the flow \cite{castaing1989} and/or the nature of the BLs \cite{siggia1990} have been proposed to explain the $2/7$ scaling. For extremely large $Ra$, however, $Nu$ is predicted to become independent of the molecular properties of the fluid, and hence the boundary layers, and heat transport is achieved solely by convection \cite{kraichnan1962, spiegel1971}. In this so-called ``ultimate regime", $Nu \sim Ra^{1/2}$ \cite{spiegel1971}.  Finally, we note that a means of examining the various ``crossovers'' in the $Ra-Pr$ plane has been proposed \cite{stevens2013}.  

Taking a different approach, Howard \cite{howard1963} sought to determine upper bounds on $Nu$ using a variational formulation 
with incompressibility as one of the constraints on the statistically stationary flow. 
When a single horizontal wavenumber dominates the flow, he found an upper bound of $Nu - 1 = \left(Ra/248\right)^{3/8}$.
Kerswell \cite{Kerswell:1998} and Hassanzadeh \emph{et al.} \cite{hassanzadeh2014} (and references therein) provide a detailed discussion of this approach. A recent variational study of the two-dimensional problem by Whitehead \& Doering \cite{whitehead2011} has shown that the thermal BLs (TBLs) do play a role in limiting the heat flux in the ``ultimate regime'', even when there are no momentum boundary layers (free-slip conditions were used). Hence, 
the nature of the interaction between the BLs and the core flow plays a central role in turbulent Rayleigh-B\'enard convection. 

This interaction can be probed either by manipulating the boundary geometry itself or by introducing inhomogeneous temperature boundary conditions \cite{biferale2014}. The former can be achieved by corrugating one or both horizontal boundaries, although we argue here that an asymmetric geometry provides unique insights. Earlier studies on convection over rough surfaces were motivated by the need for a better understanding of the role of BLs in the high $Ra$ regime \cite{shen1996, du2000}. The geometry used by Shen \emph{et al.} \cite{shen1996} and Du \& Tong \cite{du2000} consisted of a cylindrical cell with rough top and bottom boundaries made of pyramidal elements. The ratio of wavelength ($\lambda^*$) to amplitude of roughness ($h^*$) was fixed at $2$ ($\gamma \equiv \lambda^*/h^* = 2$). When the thickness of the thermal boundary layer was smaller than $h^*$, the pre-factor of the scaling relation $Nu = A \times Ra^{\beta}$ increased. There was also an increase in the plume production with an enhanced detachment near the tip of the pyramids. Similar observations were made by Ciliberto \& Laroche \cite{ciliberto1999} in which the rough surfaces were prepared by gluing glass spheres to the bottom plate and coating the surface with a thermally conductive paint. 

The experiments of Qiu \emph{et al.} \cite{qiu2005} and the direct numerical simulations of Stringano \emph{et al.} \cite{verzicco2006} have shown that $\beta$ changes for periodic roughness, with $\beta = 0.35$ in experiments and $\beta = 0.37$ in the simulations. Wei \emph{et al.}'s \cite{wei2014} experiments with different combinations of smooth and rough surfaces (with pyramidal elements of $\gamma = 2$) at the top and bottom of the cell revealed that for: (a) both surfaces rough: $\beta = 0.35 \pm 0.01$, (b) only the top surface rough: $\beta = 0.32 \pm 0.01$ and (c) only the bottom surface rough: $\beta = 0.29 \pm 0.01$. However, Dirichlet (Neumann) conditions are applied on the top (bottom) surface.

Similar studies have been carried out using rough surfaces made of rectangular elements \cite{shishkina2011, tisserand2011, salort2014}. Tisserand \emph{et al.} \cite{tisserand2011} investigated the effects of a bottom rough boundary on the heat transport near a planar top boundary by analyzing $Nu(Ra)$ separately at the top and bottom boundary.  They found that the smooth top boundary is not influenced by the effects of the bottom rough boundary.

Surface roughness has also been used in attempts to reach the ultimate regime at $Ra$ smaller  than predicted by the theory of Kraichnan \cite{kraichnan1962}. Roche \emph{et al.} \cite{roche2001} used a cylindrical cell with an interior entirely covered with V-shaped grooves to trigger a transition to turbulence in the BLs, and they reported $Nu \sim Ra^{1/2}$ for $Ra > 2 \times 10^{12}$. Detailed accounts of the developments spanning various periods with a variety of perspectives can be found in a number of reviews \cite{siggia1994, ahlers2009, schumacher2012}.

Despite differences in characteristics of turbulent flows in two and three dimensions (3D) (e.g., \cite{boffetta2012}), numerical studies of 2D Rayleigh-B\'enard  convection for large $Ra$ and $Pr \ge 1$ have yielded $Nu(Ra)$ surprisingly close to those from experiments, the differences being principally in the pre-factors \cite{doering2009, deluca1990, waleffephys2015}.
Clearly, this correspondence provides an essential role for well resolved 2D simulations to probe the properties of the key components (BLs, plumes, core flow) of convective flow and to relate them to the 3D dynamics for $Pr \ge 1$ \cite{schmalzl2004}. However, we note that the pre-factors in 3D are larger and hence so too are the values of $Nu$ \cite{van2013}.

Here, we describe quantitative studies of the effects of sinusoidal roughness of the upper boundary, with fixed amplitude and varying wavelength, on $\beta$ in two-dimensional Rayleigh-B\'enard convection using highly resolved Lattice Boltzmann method (LBM) numerical simulations.  We use this roughness to systematically probe the coupling between the BLs and the core flow and hence the resulting changes in the heat transport. 
We find that the heat transport is maximized at a dimensionless wavelength $\lambda \equiv \lambda_{\text{max}} \approx (2 \pi)^{-1}$, with $\beta = 0.359$, and decays to the planar value ($\beta \approx 0.28$) in both the large ($\lambda \gg \lambda_{\text{max}}$) and small ($\lambda \ll \lambda_{\text{max}}$) wavelength limits.  This maximum originates in the nature of the coupling between the boundary layer and the bulk flow.

\section{Governing Equations and Numerical Method}

\begin{figure}
\centering
\includegraphics[trim = 0 180 0 50, clip, width = 0.9\linewidth]{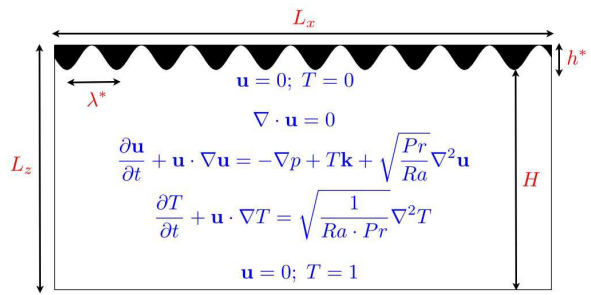}
\caption{Two-dimensional rectangular cell with $\Gamma = 2$. The cell is periodic along the horizontal direction. No-slip and Dirichlet conditions are enforced at the bottom plate and on the rough upper surface.} 
\label{fig:domain}
\vspace{-0.15 in}
\end{figure}

We describe thermal convection with the Oberbeck--Boussinesq equations \cite{chandra2013}. We non-dimensionalize them by choosing the height of the cell, $L_z$, as the length scale, the temperature difference across this height, $\Delta T$, as the temperature scale, $\mathrm{U_0} = \sqrt{g \alpha \Delta T L_z}$ as the velocity scale, where $g$ is the acceleration due to gravity and $\alpha$ is the thermal expansion coefficient of the fluid, and $t_0 = L_z/U_0$ as the time scale. The dimensionless equations of motion and boundary conditions are shown in figure \ref{fig:domain}.
Here, ${\mathbf u({\mathbf x}, t)}$, $T({\mathbf x}, t)$ and $p({\mathbf x}, t)$ are the velocity, temperature and pressure fields respectively, ${\bf k}$ is the unit vector along the vertical axis, $Ra = g \alpha \Delta T L_z^3/\nu \kappa$, and $Pr = \nu/\kappa$. The heat transfer rate in terms of $Nu$ can be obtained from $Nu = -\left<\partial_z T\right>_{z=0}$, where $\left<...\right>$ denotes a horizontal and temporal average taken after the statistically steady state has been reached. 

The governing equations are solved using the LBM \cite{benzi1992, chen1998, succi2001} with separate distribution functions for the momentum and temperature fields \cite{shan1997, guo2002}. For planar geometries with horizontal periodicity, one can solve the macroscopic equations numerically using spectral methods \cite{orszag1971}, which provide a natural way to cluster grid points near the boundaries where steep gradients in temperature result for $Ra \gg 1$ and $Pr = O(1)$, and hence where resolution is important. However, employing spectral methods for rough geometries is both technically challenging and computationally demanding, whereas the LBM can handle rough geometries naturally \cite{chen1998, succi2001}. Given our focus on the interaction between the boundary layer, which is `perturbed' by the imposed roughness, and the core flow, it is advantageous to use the LBM. For all the simulations reported here, a mid-grid bounce-back condition is used to enforce the no-slip condition at the solid boundaries, and Dirichlet conditions for temperature \cite{succi2001}. Periodic boundary conditions are used along the horizontal.

\begin{figure}
\centering
\includegraphics[trim = 0 0 0 0, clip, width = 0.9\linewidth]{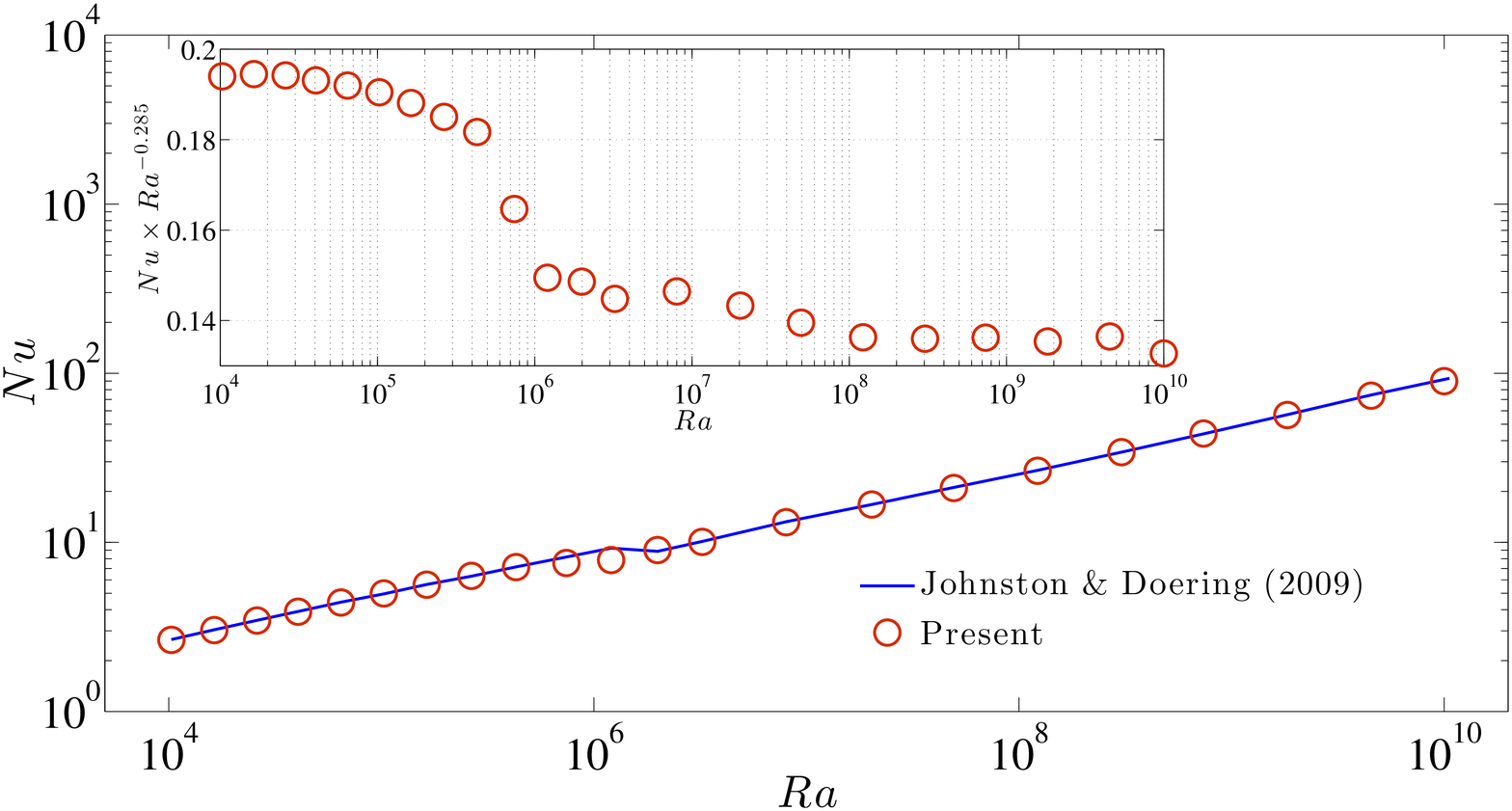}
\caption{Comparison with Johnston \& Doering \cite{doering2009} for $\Gamma = 2$ and $Pr=1$. The inset shows the compensated plot for the data. The highest eight $Ra$ data points can be fit with $Nu = 0.138 \times Ra^{0.285}$.}
\label{fig:jd}
\vspace{-0.15 in}
\end{figure}

The code developed has been parallelized using the Message Passing Interface system, and extensively tested by reproducing a range of classical results from different flows  \cite{lipps1976, clever1974, simakin1984, doering2009}. The simulations of Johnston \& Doering \cite{doering2009} with planar upper and lower plates constitute the most relevant Rayleigh-B\'enard test comparison.  They used a Fourier-Chebyshev spectral method with at least $8$ grid points inside the thermal boundary layers (TBLs). The aspect ratio $\Gamma$ -- ratio of cell width ($L_x$) to height ($L_z$) -- is fixed at $2$. We use a uniform grid with at least $8$ grid points in the TBL, ensuring very high resolution throughout the domain. Figure \ref{fig:jd} shows the comparison of our $Nu(Ra)$ with that of Johnston \& Doering \cite{doering2009} for $Pr = 1$. We recover their results for $Ra = \left[10^4, 10^{10}\right]$, and their fit of the highest eight $Ra$ data points with $Nu = 0.138 \times Ra^{0.285}$.  We note that this is also remarkably close to $Nu = 0.172 \times Ra^{0.285}$ obtained experimentally by Urban \emph{et al.} \cite{urban2011} using cryogenic \ce{^{4}He} gas in a cylindrical cell of $\Gamma = 1$ for $Ra = \left[7.2 \times 10^6, 10^{11}\right]$ and $Pr \approx 1$.

\section{Results}
\subsection{Simulations}
We now replace the upper planar boundary by a sinusoidal surface of wavelength $\lambda^*$ and amplitude $h^*$ (figure \ref{fig:domain}). For all the simulations discussed here, $\Gamma = 2$, $h \equiv h^*/L_z = 0.1$, and $Pr = 1$. We consider: $\lambda \equiv \lambda^*/L_z = 0.03, 0.05, 0.1, 0.125, 0.154, 0.2, 0.286, 0.4, 0.5, 0.67$, and  $1$. The characteristic vertical length scale used to define $U_0$, $t_0$ and $Ra$ is $H = (L_z - h^*)$. Simulations were for performed for $Ra = \left[10^6, 2.5\times 10^9\right]$ for all $\lambda$. To give an example of the resolutions used: For the smallest $\lambda$ ($= 0.03$) and highest $Ra$ ($= 2.5 \times 10^9$) the number of grid points we used along the horizontal and vertical directions was $N_x = 2500$ and $N_z = 1250$.  The wavelength and amplitude of each roughness element was resolved using $38$ and $125$ grid points respectively. With $t_0$ the turnover time, the total run time for the highest $Ra$ cases was {\em no less than} $215 t_0$, and data for statistics were collected only after $t = 108 t_0$. The run times were longer for smaller $Ra$. Grid independence was confirmed from simulations with $Ra = 10^8$ and $\lambda = 0.05$ for two grids: (a) $N_x = 1800, N_z = 900$ and (b) $N_x = 840, N_z = 420$. The difference in $Nu$ for the two simulations was $1.5$ \% and the averaged temperature profiles were indistinguishable. 

\begin{figure}
    \centering
    \begin{subfigure}
       \centering
        \includegraphics[trim = 0 0 0 0, clip, width = 0.8\linewidth]{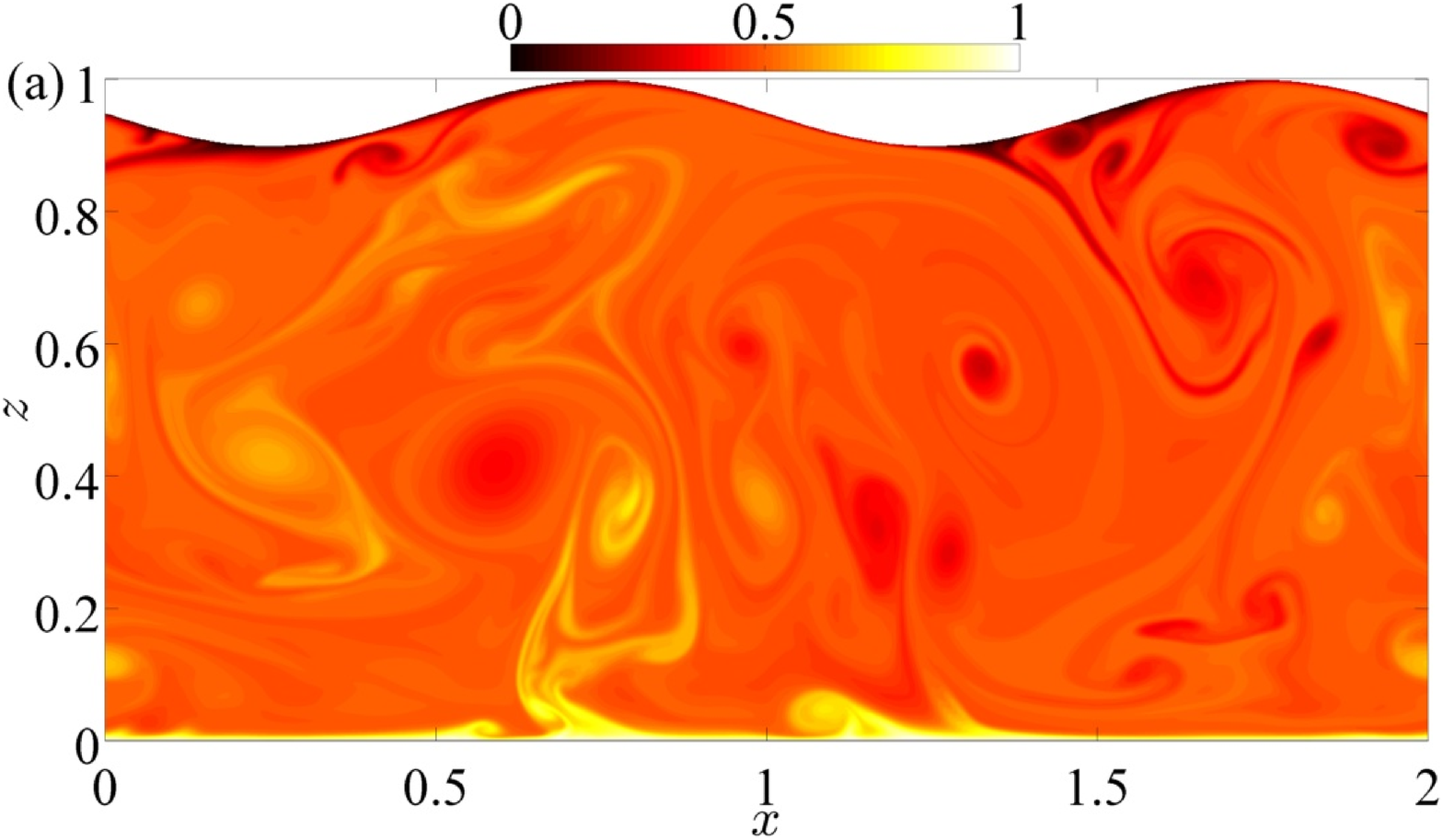}
    \end{subfigure}
    
    \begin{subfigure}
       \centering
        \includegraphics[trim = 0 0 0 0, clip, width = 0.8\linewidth]{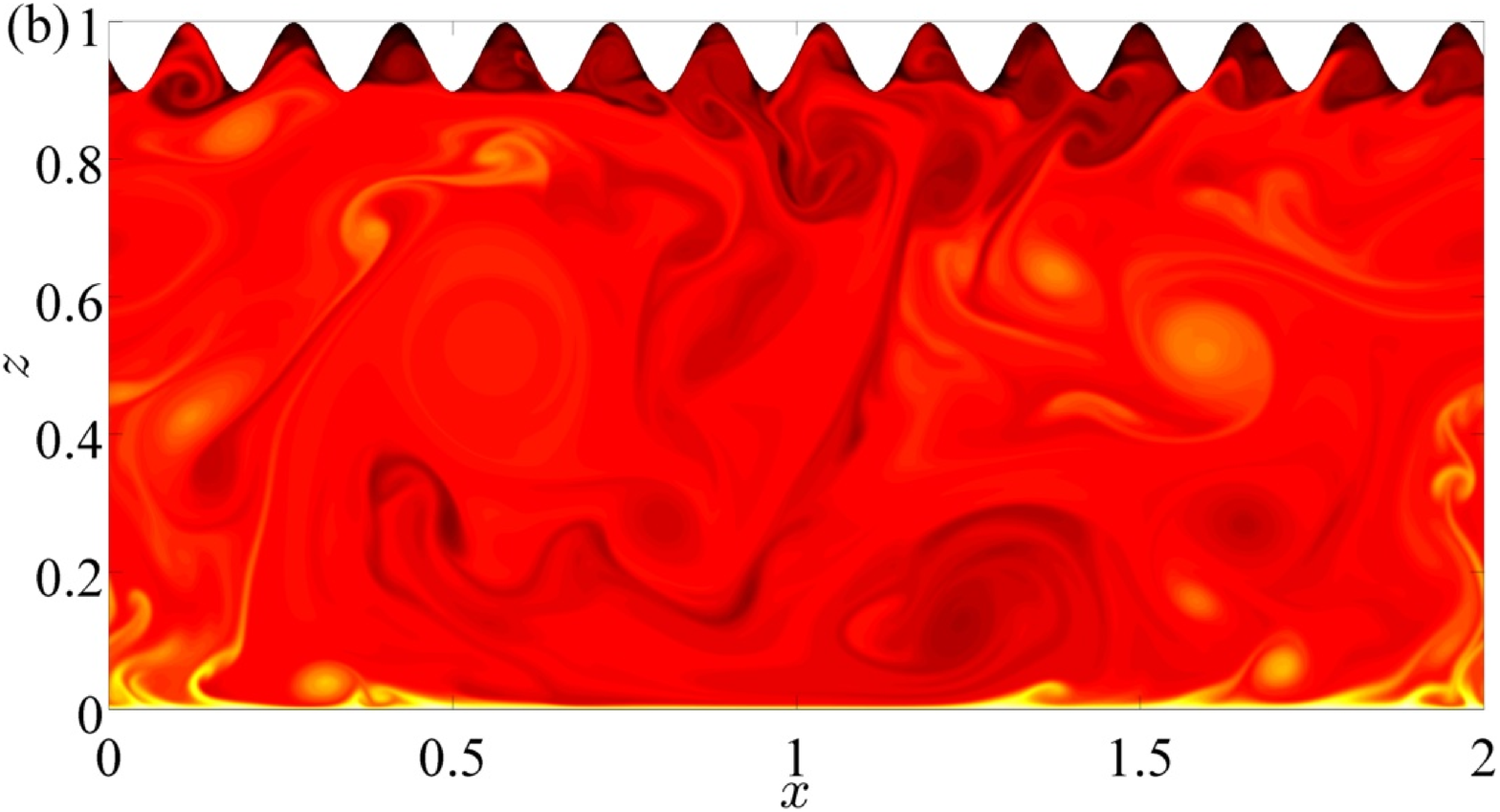}
          \end{subfigure}
          
          \begin{subfigure}
       \centering
        \includegraphics[trim = 0 0 0 0, clip, width = 0.8\linewidth]{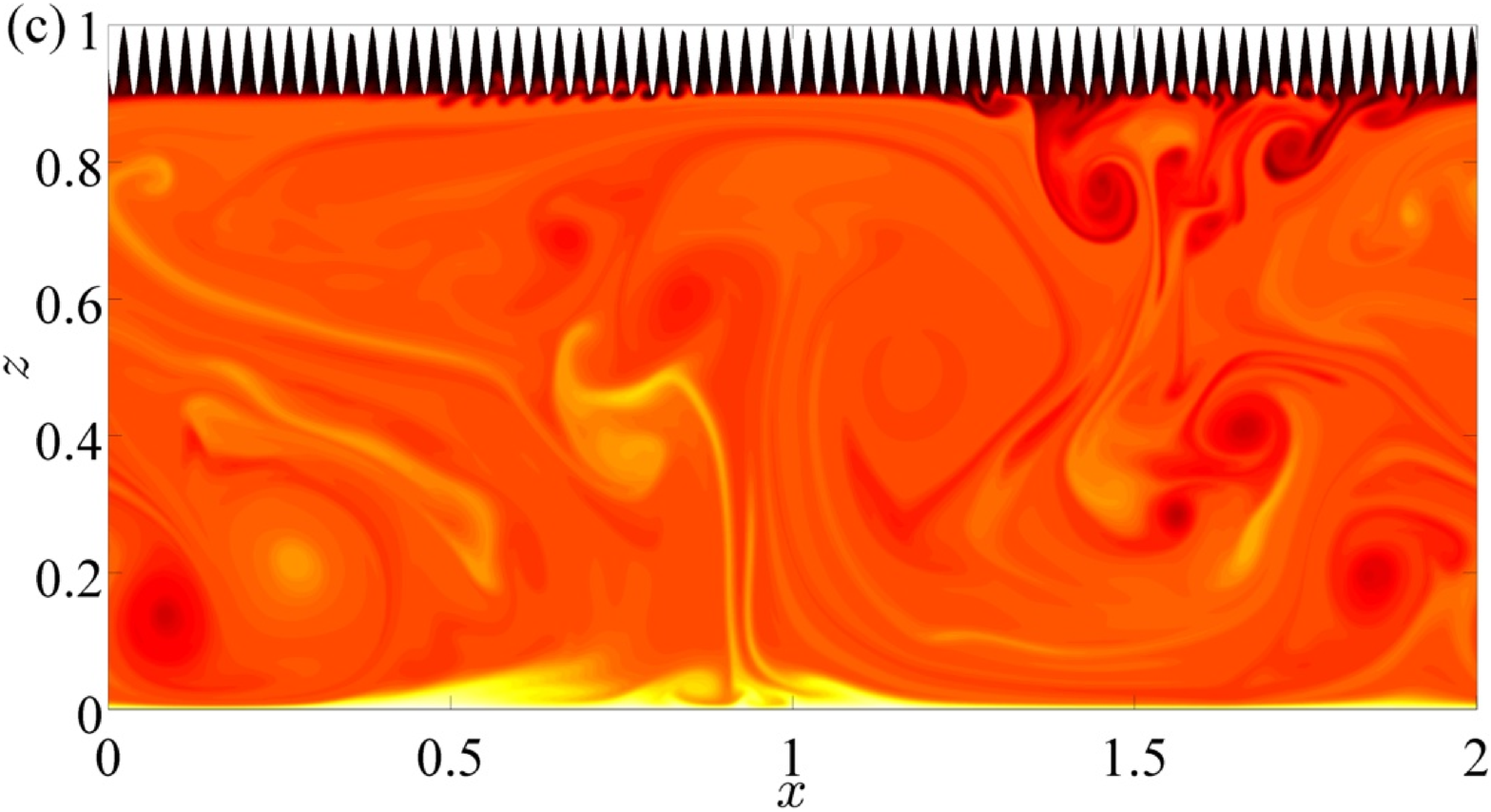}
          \end{subfigure}

\caption{Temperature field for $Ra = 10^9$ and: (a) $\lambda = 1.0$, (b) $\lambda = 0.154$ and (c) $\lambda = 0.03$. The probability density of thermal fluctuations as a function of depth (not shown) and the associated skewness, show the number of plumes produced at the rough surface is maximal for $\lambda = 0.154$.}
\label{fig:temp_rough}
\vspace{-0.15 in}
\end{figure}

Figure \ref{fig:temp_rough} shows temperature fields for $Ra = 10^9$ and $\lambda = 1.0, 0.154$ and $0.03$.  Analysis of the data from all runs reveals that the number of plumes produced at the rough surface is a maximum for $\lambda = 0.154 \equiv \lambda_{\text{max}}$, demonstrating an enhanced interaction between the boundary layer and the core flow relative to the long- and short-wavelength cases.  Namely, as $\lambda$ increases or decreases relative to $\lambda_{\text{max}}$, this interaction weakens leading to a lower production of plumes. This effect can also be seen in the average temperature field within the well-mixed core region $\langle T \rangle$. For $\lambda = 0.03$ and $1.0$ we have $\langle T \rangle \approx 0.5$ as in the planar case, but for $\lambda = \lambda_{\text{max}}$ we find $\langle T \rangle \approx 0.4$. This clearly shows the effect of the roughness element wavelength on the dynamics of the cold plumes that are released from the element tips. While previous experiments \cite{shen1996, du2000, qiu2005, wei2014} and numerical simulations \cite{verzicco2006} have reported enhanced plume production, here we find a {\em wavelength dependence}, which exhibits a maximum plume production at a particular wavelength $\lambda_{\text{max}}$.

\begin{figure}
\centering
\includegraphics[trim = 50 175 100 200, clip, width = 0.9\linewidth]{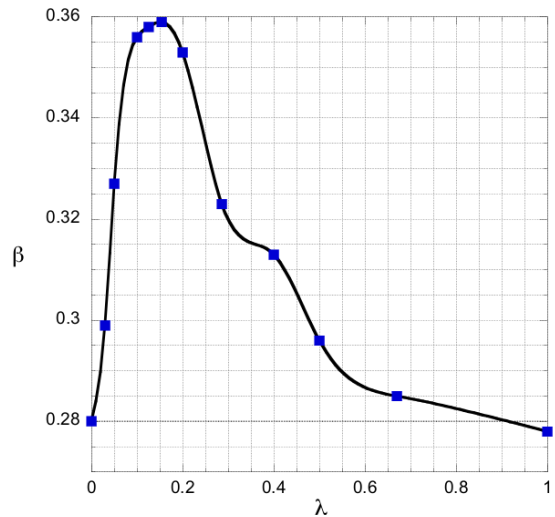}
\caption{Scaling exponent ($\beta$) as a function of roughness wavelength ($\lambda$). The solid squares represent data from simulations; $\beta$ is taken to be the planar case  for $\lambda = 0$ due to continuation of the curve; the solid line is a spline. Here, $\beta = 0.359$ at $\lambda = 0.154 \equiv \lambda_{\text{max}}$ is the maximum.  Alternatively, rather than a spline we could fit $\lambda < \lambda_{\text{max}}$ with a Gaussian and $\lambda_{\text{max}} \le \lambda \le 1$, with an exponential: $\beta = 0.116 \times \mathrm{e}^{-4.46 \times \lambda} + 0.280$.}
\label{fig:beta}
\vspace{-0.15 in}
\end{figure}

Because plumes are the conveyors of heat from one BL to another \cite{howard1966}, this change in the dynamics has a direct effect on $\beta$. Figure \ref{fig:beta} shows $\beta$ as a function of $\lambda$. For each $\lambda$, $\beta$ was determined from a linear least squares fit to the $Nu-1$ vs. $Ra$ data for $Ra = \left[4 \times 10^6, 2.5\times 10^9\right]$. Clearly, $\beta$ is maximal ($=0.359$) at $\lambda = \lambda_{\text{max}}\approx1/{2 \pi}$, 
and when $\lambda \ll \lambda_{\text{max}}$ or $\lambda \gg \lambda_{\text{max}}$ we recover the planar boundary value of $\beta$.  A dimensional argument 
providing $\lambda_{\text{max}}$ inscribes a negatively buoyant parcel of vertical ($h^{*}$) to horizontal ($2 \pi h^{*}$) aspect ratio to the upper sinusoidal grooves.  However, there is a more complex $\lambda$ dependence of the flux as described next. 

\begin{figure}
\centering
\includegraphics[trim = 0 0 0 0, clip, width = 0.9\linewidth]{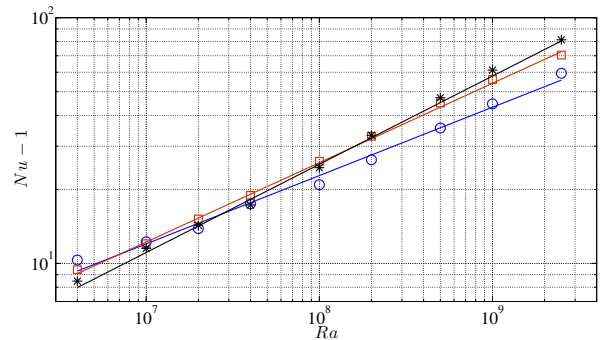}
\caption{Linear least square fits for $Nu-1$ vs. $Ra$ data for different $\lambda$. (a) $\lambda = 1.0$ (blue): $\bigcirc$: simulation data; solid line: $Nu-1 = 0.136 \times Ra^{0.278}$. (b) $\lambda = 0.286$ (red): $\Box$: simulation data; solid line: $Nu-1 = 0.067 \times Ra^{0.323}$. (c) $\lambda = 0.154$ (purple): $\ast$: simulation data; solid line: $Nu-1 = 0.034 \times Ra^{0.359}$.}
\label{fig:fits}
\vspace{-0.15 in}
\end{figure}

\begin{figure}
\centering
\includegraphics[trim = 0 0 0 0, clip, width = 0.9\linewidth]{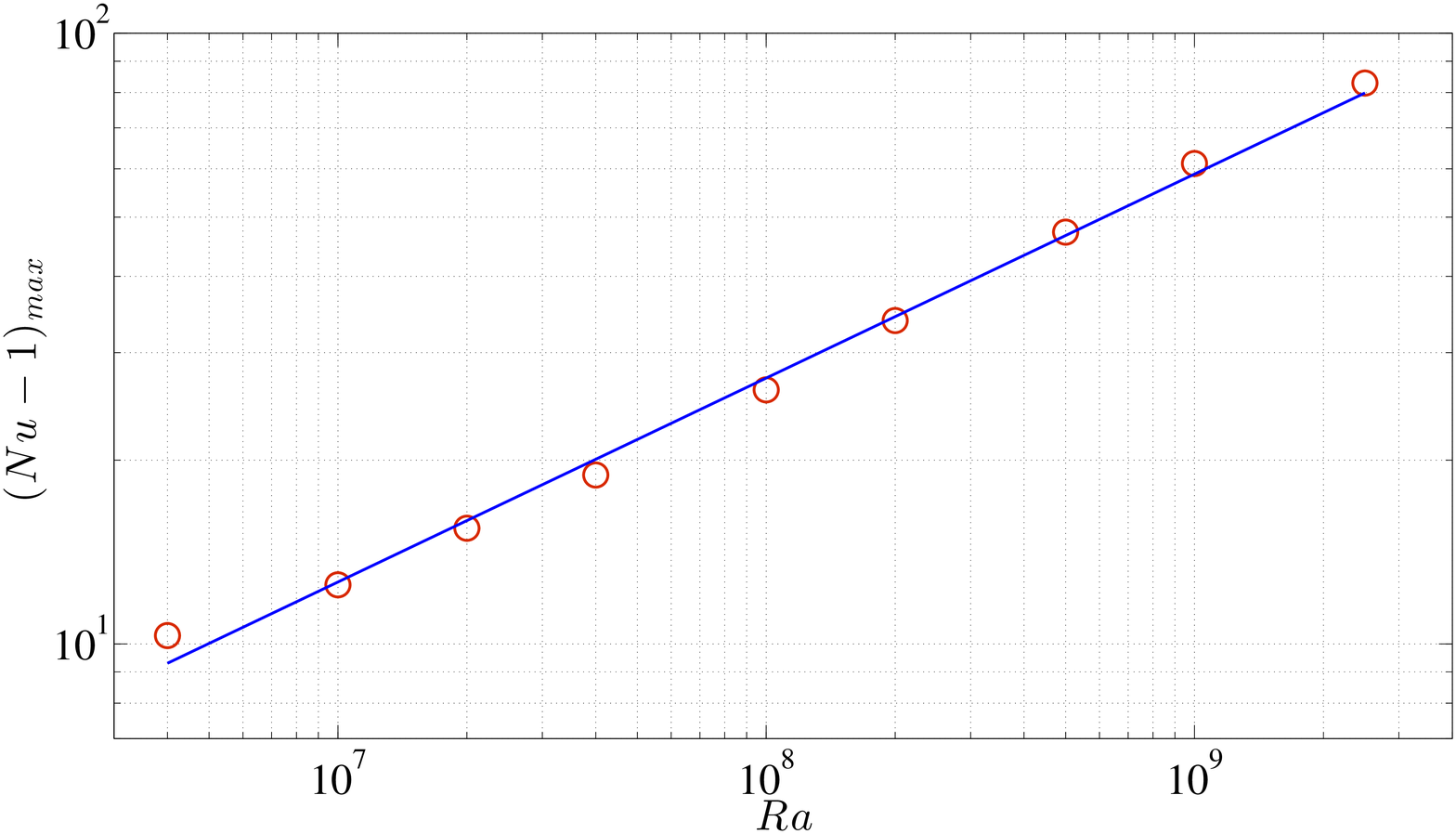}
\caption{Scaling relation for $\left(Nu-1\right)_{max}$ vs. $Ra$. For a given $Ra$, $\left(Nu-1\right)_{max}$ is the maximum value of $\left(Nu-1\right)$ among all $\lambda$ considered. Circles in the figure are data from simulations, and the solid line is the fit $Nu-1 = 0.058 \times Ra^{0.334}$ obtained from a linear least squares fit.}
\label{fig:nu-max}
\vspace{-0.15 in}
\end{figure}

Although figure \ref{fig:beta} most clearly demonstrates the main point, figure \ref{fig:fits} shows that the higher flux is dominated by the large $Ra$ contributions, where the heat flux increases as wavelength decreases.  However, at lower $Ra$ the heat flux decreases as wavelength decreases. This wavelength dependence is demonstrated further in figure \ref{fig:nu-max} where we plot the results of the following analysis.  For a given $Ra$ we determine the maximum value of $Nu-1$ among all of the $\lambda$ considered and we define this as $\left(Nu-1\right)_{max}$.  As shown in figure \ref{fig:fits} the heat fluxes are larger at small (large) $\lambda$ for large (small) $Ra$ and hence $\left(Nu-1\right)_{max}$ averages over these two competing wavelength dependent high and low flux behaviors giving a $\beta = 0.334$.  This is the average over the large and small $\lambda$ values seen in figure \ref{fig:beta}.   

Finally, we comment on two experiments.  Firstly,  as we discussed above, surface roughness has been used in hopes of reaching the ultimate regime at $Ra$ smaller than predicted by the theory of Kraichnan \cite{kraichnan1962}.  We note that the effect of roughness enhancing transport is seen in figure \ref{fig:nu-max} where we find $Nu-1 = 0.058 \times Ra^{0.334}$ for $Ra$ less than $10^{10}$, which is very nearly that found by Urban \emph{et al.} \cite{urban2011} for $Ra$ greater than $10^{11}$ for planar surfaces.  Secondly, we highlight the results for $\lambda = 0.2$ (or $\gamma = 2$) because of the close correspondence to one of the geometries used by Wei \emph{et al.} \cite{wei2014} in their experiments. In this case we obtain $Nu = 0.052 \times Ra^{0.339}$ and they obtain $Nu = 0.099 \times Ra^{0.32 \pm 0.01}$.  The agreement is remarkable given that they used a cylindrical cell of $\Gamma = 1$ with a rough top plate made of pyramidal elements of $\gamma = 2$, and highlights the importance of a systematic experimental studies in which the geometry of one surface is changed. 

\subsection{Scaling Arguments}

We now propose a simple scaling argument for the maximal  exponent $\beta$ that has been observed here. We assume that: ($1$) The flow field is dominated by plumes. ($2$) The cold plumes that are generated at the rough boundary are due to negative buoyancy. ($3$) The flow in the core region is dissipation free, and hence the energy is dissipated only in the boundary layers. By way of analogy, we appeal to the Kolmogorov picture of 3-D turbulence to understand the kinetic energy transfer from the cold plumes to the momentum boundary layer (MBL) on the other side of the cell, which acts as a sink. A similar analogy has been used in the study of wall-bounded turbulent flows \cite{sreenivasan1989}. Negative buoyancy `injects' energy into the cold plumes at the upper rough boundary, which then travel through the core region into the MBL on the opposite side, where they are dissipated. The velocity scale of the plumes is $u_p \sim \sqrt{g \alpha \Delta T h^*/2} = \sqrt{g \alpha \Delta T \lambda^*/2\gamma}$, and the velocity scale in the MBL is $u_{bl} \sim \nu/\delta_v^*$, where $\delta_v^*$ is the dimensional thickness of the MBL. The scale for $u_{bl}$ implies that inertial and viscous effects are of similar order close to the boundaries \cite{schlichting2000}; in other words $Re = O(1)$ in the BLs \cite{zocchi1990}. The length scale associated with the plumes is $\lambda^*$. Hence, the rate of injection of energy per unit mass is $\epsilon_i \sim u_p^3/\lambda^*$, and the energy dissipation rate per unit mass is $\epsilon_d \sim \nu (u_{bl}/\delta_v^*)^2 = \nu^3/\delta_v^{*4}$. In the statistically steady state we must have $\epsilon_i \approx \epsilon_d$. After some manipulation we obtain:
\begin{equation}
\frac{H}{\delta_v^*} \approx \left(\frac{1}{2 \gamma}\right)^{3/8} \left(\frac{\lambda}{H}\right)^{1/8} Pr^{-3/8} Ra^{3/8},
\label{eqn:scaling1}
\end{equation}
and noting that $\delta_v^*/\delta_T^* = \sqrt{Pr}$, where $\delta_T^*$ is the dimensional thickness of the TBL, and $\delta_T^* = H/2 Nu$ \cite{howard1966}, we arrive at 
\begin{equation}
Nu \approx \left(\frac{1}{2 \gamma}\right)^{3/8} \left(\frac{\lambda^*}{256 H}\right)^{1/8} Pr^{1/8} Ra^{3/8}.
\label{eqn:scaling2}
\end{equation}
The theoretical exponent ($\beta = 0.375$) is close to that obtained from our numerical simulations ($\beta = 0.359$), and our scaling argument embodies the dynamics of an optimal--dissipation free core--flow interacting with the BLs.  Hence, the small difference in $\beta$ is due to (a) the assumption that the core region is dissipation free (an obvious oversimplification) and (b) finite size effects in the simulations. Finally, although we are aware that the relation between $\delta_v^*$ and $\delta_T^*$ could perhaps be refined, it is not an essential issue for our argument given the rather weak $Pr$ dependence found here in equation \ref{eqn:scaling2}.

\vspace{-0.1in}
\section{Conclusions}

We have investigated the effects of an upper rough boundary with fixed amplitude $h$ and varying wavelength $\lambda$ on turbulent convection in two dimensions over a wide range of Rayleigh number $Ra$.   In this manner, we systematically manipulated the interaction between the boundary layer and interior/core flows.  We have shown that there is a wavelength $\lambda_{\text{max}}\approx1/{2 \pi}$ that maximizes heat transport, whereas for small ($\lambda \ll \lambda_{\text{max}}$) and large ($\lambda \gg \lambda_{\text{max}}$ ) wavelengths the planar case (lower flux) results are recovered.  

This wavelength dependence of the heat flux is reflected in the non-monotonic behavior of $\beta(\lambda)$ in the Nusselt-Rayleigh scaling relation, $Nu - 1 \propto Ra^\beta$ shown in figure \ref{fig:beta}.  For $\beta(\lambda_{\text{max}})$, the boundary-layer/core-flow interaction is enhanced by an increase in the number of plumes produced along the roughness elements and their direct injection from the tips into the core flow, circumventing an intermediate transition region.  The effect of the enhanced upper surface plume injection is to decrease the average core temperature $\langle T \rangle$ relative to $\lambda \ll \lambda_{\text{max}}$ and $\lambda \gg \lambda_{\text{max}}$.  

The $Nu \propto Ra^\beta$ relation obtained for $\lambda = 0.2$ ($\gamma = 2$) is in agreement with the recent experiments of Wei \emph{et al.} \cite{wei2014}, highlighting the importance of a systematic experimental study in which the geometry of one surface is changed. 

A simple scaling argument describing the dynamics of a maximal flux has been proposed, and it is in good agreement with the simulation results.  This prompts us to speculate on the relation between $\beta(\lambda_{\text{max}})$ and variational approaches that seek maximal fluxes in the planar case for single wavenumber flows, such as Howard's \cite{howard1963} treatment of optimal heat flux and Doering and Constantin's \cite{Doering:1996} upper bound using background test fields. In both cases, for fixed $Pr$,  $\beta = 3/8$ which is similar to $\beta = 0.359$ obtained here numerically.  Clearly, detailed theory, numerics and experimentation is necessary for a firm understanding of this correspondence.  Furthermore, we note that the robust $\beta = 0.5$ upper bound scaling holds not only for flat no-slip boundaries, but also for both one- and two smoothly modulated boundaries (Goluskin \& Doering, pers. comm.).  In our ongoing but preliminary simulations over $Ra = \left[4 \times 10^6, 2 \times 10^9,\right]$ for both upper and lower boundaries having $\lambda = 0.154$ ($\lambda = 0.2$), we find a pre-factor of 0.0055 (0.0091) and $\beta = 0.471$ ($\beta = 0.442$).  

Finally, given the fact that outside of the laboratory setting the boundaries of convecting fluids are rarely uniform, the results presented here have important implications for turbulent transport in astrophysical \cite{mitra2013}, engineering \cite{schumacher2012} and geophysical \cite{bercovici2007} settings.  

\acknowledgments

The authors acknowledge the support of the University of Oxford and Yale University, and the facilities and staff of the Yale University Faculty of Arts and Sciences High Performance Computing Center. ST acknowledges a NASA Graduate Research Fellowship and Sumesh P. T. for helpful discussions.  JSW acknowledges the Swedish Research Council and a Royal Society Wolfson Research Merit Award for support.    This work was completed at the 2015 Geophysical Fluid Dynamics Summer Study Program at the Woods Hole Oceanographic Institution, which is supported by the National Science Foundation and the Office of Naval Research.  We thank many of the staff for feedback, particularly C.R. Doering \& D. Goluskin. 

\bibliography{EPL}

\end{document}